\documentclass[pra,twocolumn,nofootinbib]{revtex4}

\usepackage{graphicx}
\usepackage{dcolumn}
\usepackage{bm}
\usepackage{amssymb}
\usepackage{amsmath}
\usepackage{amsfonts}
\usepackage[]{graphicx}
\def\beq{\begin{eqnarray}}
\def\eeq{\end{eqnarray}}

\begin{document}
\bibliographystyle{prsty}

\title{ Comments on `Comment on Aur\'elien Drezet's defense of relational quantum mechanics'  by Jay  Lawrence, Marcin Markiewicz and Marek \'{Z}ukowski. }

\begin{abstract}
We respond briefly to the recent comment by Jay  Lawrence, Marcin Markiewicz and Marek \'{Z}ukowski [arXiv:2210.09025  and Found. Phys. \textbf{54}, 45 (2024)] regarding our work defending RQM against their previous assessment. We refute the analysis proposed by the authors and rephrase our previous study in order to clarify the remaining ambiguities in our rebuttal.
\end{abstract}
\author{Aur\'elien Drezet}
\affiliation{Institut N\'eel, UPR 2940, CNRS-Universit\'e Joseph Fourier, 25, rue des Martyrs, 38000 Grenoble, France}

\date{\today}
\maketitle

\section{Introduction}\label{sec1}
 
 In a recent paper published elsewhere \cite{1}, authors Jay Lawrence, Marcin Markiewicz, and Marek \'{Z}ukowski (hereinafter LMZ) presented a technical argument in an attempt to refute the Relational interpretation of quantum mechanics (RQM) developed by Carlo Rovelli since 1996. \cite{Rovelli1996,Rovellibook2021,RovelliFP,Rovellihandbook} We stress that RQM is to the standard Copenhagen interpretation what  special relativity is to Newtonian mechanics. \footnote{It is important to note that RQM, using a Heisenberg-type separation between observed and observer systems, is far less vague than the Qbist approach, which involves the notion of a conscious agent (the precise nature of which is very poorly defined), or the Bohrian approach, which assumes a kind of `irreversible' thermodynamical amplification during a measurement process, again escaping any precise definition (i.e., assuming a vague notion of wave function collapse). In this sense, RQM appears to be the most precise and unambiguous formulation of Heisenberg's original ideas. We also stress that RQM is motivated by Rovelli's research in loop quantum  gravity where the Copenhagen interpretation doesn't make sense.}  In a note published in this journal \cite{2}, however, the present author demonstrated that the so-called refutation described in \cite{1} in fact contained significant conceptual errors invalidating it in its entirety. As is often the case in quantum mechanics, many purported proofs of impossibilities in fact demonstrate the existence of prejudices that limit and sometimes invalidate such proofs (e.g., the infamous Von Neumann no-go theorem).  The attempt published in \cite{1} is a striking demonstration of this fact, as is the counter-analysis \cite{2} refuting its prejudices.  
LMZ have now presented in the same journal a comment \cite{3} as an attempt to refute the argument described in \cite{2}, and the purpose of this note is to demonstrate that their new analysis again contains elementary misunderstandings concerning contextuality in both RQM and (more surprisingly) the standard Copenhagen interpretation. 

Curiously, \cite{3} does not focus on the mathematical argument of \cite{2} defending RQM against the attack described in \cite{1} (the authors even ironically acknowledge that the calculations in \cite{2} are correct).  Instead, LMZ claim to have found hypothetical flaws in the understanding of RQM presented in \cite{2}; which, if true, would undoubtedly invalidate, if not the defense made in \cite{2} of this same theory, at least its credibility.  This is clearly a rhetoric of denigration and avoidance which, alas, does not allow them to respond to the fatal objections contained in the note \cite{2} and which in fact still invalidate their criticism of RQM published in \cite{1}.  

The claims of \cite{3} essentially concern one point \footnote{Note that \cite{3} break down their argument into several points numbered 1 to 6. We won't follow the authors here in a tedious point-by-point response, which in fact would not shed any light on the subject.}  \cite{3}) about the role of the (reduced) density matrix in RQM.  

More precisely, LMZ claim that the  present author overestimates the role of the density matrix in the mathematical formalization of RQM. For them, the density matrix is no better than the wave function.   However, the authors of \cite{3} make an elementary confusion: in the case of a quantum system composed of two subparts A and B, it's not the total density matrix $\hat{\rho}_{AB}$ associated with the overall system that matters for RQM. As explained at length in \cite{2} and \cite{4}, and in line with Rovelli's texts \cite{Rovelli1996,Rovellibook2021,RovelliFP,Rovellihandbook}, it's the reduced matrices $\hat{\rho}_{A|B}$ and respectively $\hat{\rho}_{B|A}$ obtained by plotting respectively on the degrees of freedom of observer B or A, which must play a central role in RQM. Only these reduced matrices can reflect the relational nature of the information of one subsystem in relation to the other. \\
\indent LMZ apparently haven't read our article \cite{4}, focusing on the role of the density matrix in RQM, written in response to unfounded criticism made in \cite{5} about Wigner's friends and preferred basis in RQM. Unfortunately, many of the mistakes made in \cite{5} are now being reproduced in \cite{1,3}. We strongly encourage the readers and LMZ to go through \cite{4} where a mathematical formulation of the density matrix reading of RQM is sketched. \\
\indent In this context we stress that tracing over the degrees of freedom of the `observer' A in $\hat{\rho}_{B|A}$ implies that no self-measurement of the system A is possible in RQM. \footnote{ The self-measurement problem \cite{Breuer} is discussed in \cite{Zwirn} in the context of the RQM. Our work \cite{4}, based on the density matrix formalism, is partly motivated by the desire to solve this problem within the RQM framework.}  These reduced matrices are tools used in RQM for bookkeeping information available by the different subsystems. In other words (and in a spirit very close to the interpretation proposed by Heisenberg),  these matrices draw up an exhaustive catalog of the potentialities offered to each sub-system relative to the others. At a more technical level, partial density matrices are independent of the chosen basis, thus resolving some spurious objections to the notion of preferred basis in RQM (see the criticism in \cite{5} and our response in \cite{4}). Rovelli has always recognized the central role played by quantum decoherence in RQM (i.e., in connection with the definition of approximately stable fact  \cite{Rovelli1996,Rovellibook2021,RovelliFP,Rovellihandbook}), and partial density matrices are particularly well suited to this analysis. Note that this point also allows us to understand and resolve the so-called Wigner paradox mentioned by LMZ in \cite{3} and which was the focus of our analysis in \cite{4} in reply to \cite{5} and in full agreement with RQM.  \\
\indent Beyond these contentious points, it is important to recall the central error made by the LMZ authors in their paper \cite{1} and repeated in \cite{3} (especially in their points 4-6). In fact, in \cite{1}, they start from a particular quantum system S (an entangled spin triplet forming a GHZ state~\cite{GHZ}) and, by involving two observers A and B (also involving three qubits each), attempt to obtain a form of Wigner-friend contradiction between quantum mechanics and the RQM interpretation (previous attempts \cite{5} have already failed \cite{4}). The central point we demonstrated in \cite{2} is that LMZ actually proposes four different experiments involving the three subsystems S,A,B. By comparing these four different experimental contexts, they make counterfactual arguments that contradict the spirit of quantum mechanics (i.e. in line with the Bell and Kochen-Specker theorems). This strong contextuality of quantum mechanics has always been emphasized  by the fathers of the Copenhagen interpretation like Bohr or Heisenberg. Moreover, contextuality is also a central tenet of RQM. So it's hardly surprising that LMZ obtains contradictions with RQM. \\
\indent Although we don't wish to repeat here the derivation made in \cite{1} and the refutation presented in \cite{2}, it is nevertheless useful to briefly summarize the core of the problem in order to debunk and perhaps further clarify the so-called paradox.  
More precisely, in \cite{2} LMZ derive four relations for the four different contexts: 
\begin{eqnarray}
p^{(2)}_{SAB_1}q^{(2)}_{SAB_2}r^{(2)}_{SAB_3}=+1 \label{1}\\
p^{(2)}_{SAB_1}q^{(3)}_{SA_2}r^{(3)}_{SA_3}=-1\label{2}\\
p^{(3)}_{SA_1}q^{(2)}_{SAB_2}r^{(3)}_{SA_3}=-1\label{3}\\
p^{(3)}_{SA_1}q^{(3)}_{SA_2}r^{(2)}_{SAB_3}=-1\label{4}.
\end{eqnarray}
Here $p^{(2)}_{SAB_1}=\pm 1$ are eigenvalues associated with some specific measurements involving the first qubits of S, A and B (written hereafter   $S_1$, $A_1$ and $B_1$).  Similarly $q^{(2)}_{SAB_2}=\pm 1$ and $k^{(2)}_{SAB_3}=\pm 1$ involves respectively the second  and third qubits of S,A, B (i.e., $S_2$, $A_2$, $B_2$ and $S_3$, $A_3$, $B_3$ respectively ). Furthermore, the  eigenvalues  
$p^{(3)}_{SA_1}=\pm 1, q^{(3)}_{SA_2}=\pm 1, r^{(3)}_{SA_3}=\pm 1$ are associated with different measurements made by A on the first, second and third qubit of S and suppose that B is not interacting with S and A.\footnote{The quantum states involved in these experiments \cite{1,2} are defined by 
$|k^{(2)}\rangle_{SAB_m}:=|k^{(2)}\rangle_{SA_m}|k^{(2)}\rangle_{B_m}$  with $|k^{(2)}\rangle_{SA_m}=\frac{1}{\sqrt{2}}[|+1^{(3)}\rangle_{SA_m}+i k^{(2)}|-1^{(3)}\rangle_{SA_m}]\equiv |\textrm{sign}(k^{(2)})x\rangle_{SA_m}$ and 
$|k^{(3)}\rangle_{SA_m}:=|k^{(3)}\rangle_{S_m}|k^{(3)}\rangle_{A_m}$.  Here $m=1,2,3$ labels the 3 families  of subsystems $[S_m,A_m,B_m]$ defining Hilbert space $\mathcal{H}_{SAB_m}$, and the full quantum state belongs to the tensor product of these Hilbert spaces.}\\
\indent The first relation \ref{1}, associated with the first experimental context, supposes three measurements involving S, A and B, whereas each of the three other relations \ref{2},\ref{3} and \ref{4} involve one measurement involving S, A and B and two measurements involving only S and A.  It would be tempting to consider  $p^{(3)}_{SA_1}=\pm 1, q^{(3)}_{SA_2}=\pm 1, r^{(3)}_{SA_3}=\pm 1$ as results of measurements seen by the system A in a non-contextual way (i.e. a non contextual relative fact for A).  The relations Eqs.~\ref{2}-\ref{4}  would thus involve a mixture of relative facts for A and relative facts for B. But this would imply a paradox  since multiplying Eqs.~\ref{2}-\ref{4} clearly contradict Eq.~\ref{1}. LMZ concludes that the relative facts for A cannot be defined in a self-consistent way by B. But this is a misunderstanding for at least two reasons:

i)  Eqs.~\ref{1}-\ref{4} involves four different experimental contexts and in quantum mechanics we have generally no right to consider counterfactual reasoning mixing data taken from different contexts. This is central here since the problem involve a strongly entangled GHZ state. This problem is well known \cite{GHZ} and leads to contradictions if we think about it in a non-contextual way.

ii) For B the numbers $p^{(3)}_{SA_1}=\pm 1, q^{(3)}_{SA_2}=\pm 1, r^{(3)}_{SA_3}=\pm 1$ are not values seen by A  but instead values of S coupled to A as described by B. For example, with Eq. \ref{2} we can associate in RQM the reduced density matrix $\hat{\rho}^{(\Psi_{GHZ'})}_{SA|B}$ given Eq. 25 in \cite{2}. From this we deduce the probabilities 
\begin{eqnarray}
\mathcal{P}^{(\Psi_{GHZ'})}_{SA|B}(p^{(2)}_{SA_1|B},q^{(3)}_{SA_2|B},r^{(3)}_{SA_3|B})=\frac{1}{4}\label{5}
\end{eqnarray} assigned to the four possible alternatives verifying the constraint \begin{eqnarray}p^{(2)}_{SA_1|B}q^{(3)}_{SA_2|B}r^{(3)}_{SA_3|B}=-1. \end{eqnarray} The added label $\Psi_{GHZ'}$ reminds that this relation is obtained using a specific quantum state $|GHZ'\rangle_{SAB}$ (Eq. 20  of \cite{2}) associated with a specific physical context. Here, $p^{(2)}_{SA_1|B}$ is a result of measurement for B concerning one subsystem (i.e., the first qubit of S and the first qubit of A) but the other subsystems are not actually measured, i.e., the values $q^{(3)}_{SA_2|B}$, $r^{(3)}_{SA_3|B}$ are not actualized for B. Rather, these values $q^{(3)}_{SA_2|B}$, $r^{(3)}_{SA_3|B}$ represent potentialities or possibilities for B in agreement with Heisenberg.\footnote{We stress that if we were interested in A's perspective, we would have had to consider the reduced density matrix  $\hat{\rho}^{(\Psi_{GHZ'})}_{SB|A}$ instead of $\hat{\rho}^{(\Psi_{GHZ'})}_{SA|B}$.}

Moreover and in relation with point i, this catalog of possibilities is context dependent and for the system B it makes no sense to compare non-contextually Eqs.~\ref{1}-\ref{4} associated with different reduced density matrices.  In \cite{2} we emphasized the role of points i) and ii) which are not completely independent. Point ii) is specific to RQM and some aspects of it can be debated \cite{4,RovelliAdlam}, but point i) is actually sufficient to rule out the objections of \cite{1}.

Indeed, it is here central to consider the actual physical context and not to mix different realities in order to avoid logical contradictions.  In turn this demonstrates that far from being a refutation of RQM the results of \cite{1} (i.e., reanalysed through the good prism in \cite{2}) rather constitutes a confirmation of essential contextuality of the theory. 
    
In other words, and to sum up,  LMZ are assuming a global non-contextuality which is precisely what RQM but also standard quantum mechanics suggest to get rid of. Therefore, the fallacious analysis made in \cite{1} and \cite{3} actually misconstrues RQM and its goal, and it also miscontrues  the Copenhagen interpretation. To conclude: Many authors often criticize theories they know little about, reinterpreting them superficially in the light of their own prejudices. This can never lead to convincing arguments, and can only add confusion to an already complex situation. As we already stressed in \cite{2} and \cite{4}: We should better focus on good problems.

\end{document}